\documentclass[preprint,authoryear]{elsarticle}
\usepackage{natbib}

\begin{document}

\title{Failure of Ontological Excess Baggage as a Criterion of the Ontic Aapproaches to Quantum Theory}

\author[rvt]{Tung Ten Yong}
\ead{tytung2020@gmail.com}

\cortext[cor1]{Corresponding author}

\begin{abstract}
This article presents a discussion of the notion of quantum ontological excess baggage, first proposed by Hardy. It is argued here that this idea does not have the significance as suggested in his paper. It is shown that if this concept is properly analyzed, it fails to pose any threat to the ontic approach to quantum theory in general.  
\end{abstract}


\maketitle

\section{\label{sec:level1}Introduction}

The issue of hidden variables in quantum mechanics is essentially the problem of finding an observer independent reality behind the quantum phenomena. Hidden variables, be it deterministic or stochastic, should recover quantum mechanical predictions and experimental measurement results. Moreover, all acceptable hidden variable theories have to satisfy several no-go theorems, if they are to give rise to the familiar structure of quantum mechanics. Although these experimental and theoretical constraints do not constitute a refutation of the hidden variable theories, they impose restrictions on the form and content of any such theory. 

In the paper \cite{Hardy}, Hardy presented yet another result in reducing further the plausibility of any simplistic ontic embeddings of quantum mechanics. He showed that for any such embeddings the amount of ontological resources needed is unnecessarily large, if compared to the original, quantum resources (quantum state descriptions). In other words, the probability simplex for the set of epistemic states has infinitely many vertices, while that of the instrumental states (quantum states) is finite (this is explained in the following section). On the other hand, according to his argument, there is no such embedding overhead in the case of classical theories, eg. classical statistical mechanics. The implication of this result for the hidden variable theories is that quantum theory is very different from classical probabilistic theories in its relationship to any possible underlying ontic states. Any theory that `completes' quantum theory will have to assume a lot of superfluous quantities, infinitely many than is needed in a classical theory. This, he claimed, renders the ontic approach very unattractive.

This paper argues that such a view is too hasty. It will be shown below that if properly analyzed, infinite excess baggage also occurs in classical theories, and thus it does not at all discredit the ontic point of view. The argument is simple, but before that the relevant concepts and Hardy's result will be briefly recounted in the next section, followed by two sections on the counter arguments, and another section that discusses the correct view that should be adopted.

\section{\label{sec:level1}Relevant concepts and Hardy's result}

First of all, the instrumental state $p=(p_k)$ of an object is a mathematical entity that gives the probabilities of all the results for all possible measurements that can be carried out on the object. Knowing the correct instrumental states means knowing everything about the outcomes of any experiments. The most trivial entity that does the job is simply the list of probabilities for all these possible results. But usually there is some structure contraining or relating the set of our possible measurements (as in quantum theory), and this usually means that a shorter list of the probabilities can do the same job. So we need only to consider the lists with the smallest number of probabilities. It turns out that for a quantum object, this minimum number is equal to $N^2$ where $N$ is demension of its Hilbert space. Denote this minimum number as $K_{inst}$ (i.e. $1\leq k\leq K_{inst}$).

While the instrumental state let us predict probabilities of all outcomes in our experiment done on a physical system, we would still like to know if these states describe/contain everything there is about the reality of that physical system. The ontic states ${s_i}$ of the system describe its real state, independent of any measurements. However we may not know the real state of the system, our information/knowledge about the system is thus encapsulated by another mathematical entity - the epistemic state $P=(P_{i})$. In classical statistical mechanics, the ontic states are represented by the points in system's phase space, while the epistemic states are the probability distributions on the phase space. Now denote the total number of the system's ontic states as $K_{ontic}$. $K_{ontic}$ is thus the number of vertices of the epistemic probability simplex.

Our epistemic state reflects itself in the relation between instrumental states and ontic states: 
\begin{eqnarray} p_k =\sum \lambda^k_i P_i                                
\end{eqnarray}
where $\lambda^k_i$ denotes the probability that the ontic state $s_i$ will trigger measurement result $k$. This relation can be seen as mapping the points in epistemic probability simplex to the points in the convex set of instrumental states.

Hardy's paper then introduced the ratio 
\begin{eqnarray} \gamma:=\frac{K_{ontic}}{K_{inst}}
\end{eqnarray}
the ontological excess baggage factor, which by definition seems to be a measure of how large the ontic state space should be in order to give rise to the set of instrumental states. He showed that the factor is (a) infinite, in the quantum case; while (b) equal to unity, in the classical case\footnote{A minor point about his proof is that it actually accounts only for deterministic ontic theory. A stochastic ontic theory does not require, as is in his proof, that $A_{1}(l')$ has elements in common with $A_{1}(l)$ or $A_{2}(l)$. However this does not affect his conclusion that $\gamma$ is infinite.}. From this he concluded that any ontological completion of quantum theory requires infinitely more resources than that in the classical theories\footnote{His proof for the quantum case will not be reproduced here, as the details are irrelevant to my argument.}. 

\section{\label{sec:level1}Infinite excess baggage in the classical case}

To see the inadequacy of the above result, I will first show, using a simple example, that (b) is not unconditionally true, and that (a) does not has the significance as proposed by Hardy in his paper. This is done by demonstrating that in the classical case the ontological excess baggage factor can also be infinite:

Consider a kind of creature that possess far less knowledge than us about the molecular structure of matter. In fact, they do not know, for example, that the gas is composed of tiny molecules that contributes to the gas properties like tempetature, pressure etc. However they are able to control the total energy of the gas system, through manipulating, for example, its temperature. More precisely, they can prepare the gas with any value of total energy, and they can measure the total energy of any gas system given to him. Other than that, there are no other ways that they can manipulate the system, or know more about the system.

Therefore, the instrumental state of the gas, according to these creature, is simply $(p(H=E))$ - the list of probabilities that the total energy of the system takes the value $E$. There is not a shorter list for this because there are no further structures for the values of $H$. Next, the ontological states of the gas system, are its phase space points, as usual. These ontic states can be seen as a completion of the creature's instrumental description. Consider localized gas systems, then for any value of $H$ the system's ontic state can be any point in a finite region of phase space. 

Now what is the value of $\gamma$? Since the system can have any (positive) total energy, $K_{inst} = \infty$. Also, $K_{ontic}$ is also obviously infinite. However if careful analysis is carried out, by first discretizing the ontic state space, we obtain that for any value of $H$, there corresponds to a huge amount of ontic states. The amount of these ontic states tends to infinity when the phase space discretization goes over to continuity, for each and all values of $H$. This implies that $\gamma = \infty$.

Therefore, contrary to Hardy's argument about the classical case, in this classical world we also have infinite ontological excess baggage for the creature's instrumental states. This means that his conclusion about the quantum case (the correctness of which will be discussed in the next section) in fact does not imply a fundamentally more difficult prospect for the ontic approach to quantum theory, than for the completion of thermodynamics by the usual molecular theory (via the usual classical statistical mechanics).

Some readers may object to the above argument that I am using the wrong instrumental state here: the correct analysis should use the ontic states as the instrumental states, just like Hardy's analysis of the classical case. However this is the wrong way of vewing things. Unlike us who knows that gases are composed of molecules, these creatures cannot manipulate and do not have any knowledge about the gas ontic states. All they can control with and read out through their measurement apparatus are the values of the system's total energy $H$. By definition, instrumental state concerns only measurements that are available to them. So they should (and can only) use the probabilities of obtaining different $H$ values as the appropriate instrumental state. In general, the state space of instrumental states for a physical object (and thus $K_{inst}$) changes when we have a different class of ways of interacting with the object.

In fact, this is what constantly going on in science, for example, according to the 18th and 19th century scientists, their instrumental states for gas describes its temperature, pressure, entropy etc.; whereas for the 21st century nanoscientists their instrumental states describes the position, momentum, size etc. of the individual molecules in the gas.

\section{\label{sec:level1}The quantum case}

We can also question Hardy's argument about the quantum case. Particularly, we question his use of ontological excess baggage as a criterion for determining the plausibility of the ontic approach. Note that our issue here concerns whether excess baggage factor can serve as a legitimate criterion for choosing certain theories, not about the correctness of Hardy's argument as a logical argument that derives implications for the assumption that quantum states are the instrumental states. As a logical derivation it is undoubtedly correct. But for it to be a reasonable criterion, its truth or applicability should be independent of the details of the ontic theories.

Now, as seen from discussion above, which kind of instrumental states are the correct ones depends crucially on what we can measure on the physical object. In Hardy's argument the instrumental states are quantum states, and he shows the ontological baggage is infinite for such instrumental state. However, he did not question the role of quantum state as instrumental state in the face of our possible knowledge about the ontic states. There are ontic theories where the ontic states are measurable or distinguishable instrumentally, and thus form (entirely or part of) the instrumental states. 

As a consequence, his result is meaningful only when our knowledge about the ontic states will not require us to revise the status of quantum states as the correct instrumental state. This means that his result of infinite quantum ontic baggage only discredits (here it is granted that having an infinite value really does so, see previous section) those ontic theories with ontic states that are forever unmeasurable (lets call these theories \textbf{UOT}: unmeasurable ontic theories). But for those ontic theories that requires a different set of instrumental states than the quantum states (\textit{non}-\textbf{UOT}), his result is irrelevant.

This will then give rise to two problems. First, we don't need the concept of ontological excess baggage to tell us that \textbf{UOT} are untenable. Such ontologies are unattractive simply because they are forever hidden from observation, and yet they affect (or even give rise to) all our measurements. Second, because of its inability to deal with all ontic theories, ontological excess baggage fails as a legitimate criteria for determining the plausibility of ontic theories.

\section{\label{sec:level1}The quantum \textit{relative} ontological excess baggage}

We can view the issue in another way. Since the instrumental state of a physical system depends on observer's knowledge and his ability to measure the system, there is not a unique value of $K_{inst}$ for a physical object. What we should really have is $(K_{inst})_{level}$, the subscript `level' denoting the level of description or knowledge available to the observer about the object. For example in the above example, the creatures have $(K_{inst})_{total H}$, while human observers have $(K_{inst})_{microscopic}$ (`microscopic' denotes microscopic states - the molecular positions and momenta). 

From this we can define the quantum relative ontological excess baggage factor:
\begin{eqnarray}
\gamma_{level}:= \frac {K_{ontic}}{(K_{inst})_{level}}
\end{eqnarray}
The term `relative' highlights the fact that the factor is not absolute nor unique for a given object, but is relative to our knowledge and ability to measure it.

So for the creatures, 
\begin{eqnarray}
\gamma_{total energy} := \frac {K_{ontic}}{(K_{inst})_{total energy}} = \infty
\end{eqnarray}
While for human observers,
\begin{eqnarray}
\gamma_{microscopic} := \frac {K_{ontic}}{(K_{inst})_{microscopic}} = 1
\end{eqnarray}
In the quantum case,
\begin{eqnarray}
\gamma_{quantum} := \frac {K_{ontic}}{(K_{inst})_{quantum}} = \infty
\end{eqnarray}
Discussions in the previous sections shows that, both in the case of creature observers and that of quantum theory, having an infinite value of gamma does not imply the implausibility of ontic approaches. Now, from the prespective of relative ontological excess baggage, we find that in both of these cases $K_{ontic}$ and $(K_{inst})_{level}$ are of different levels. This suggest that the failure of Hardy's argument as a criteria is due to the fact that his ontic baggage factor, $\gamma$, involves different levels of $K$'s. 

This is in fact not difficult to understand. If the ontic states are of a different level than the instrumental states, the situation in section 4 arises - one cannot guarantee that the instrumental states will remain the same given our knowledge about ontic states. Consequently one cannot have any valid criteria based on such concept.

This urges us to propose a theorem similar to the one in Hardy's paper:
\textbf{The Quantum \textit{Relative} Ontological Excess Baggage Theorem.}\textit{Given that quantum states are the instrumental states, any interpretations/ theories of quantum theory in which the quantum states are also the ontic states are very uncomfortable. This is also true if quantum states themselves form parts or the whole of ontology.}

The proof for the first part of this theorem (when quantum states are ontic states) is essentially the same with Hardy's original theorem, but with gamma replaced by $\gamma_{quantum} := K_{ontic}/(K_{inst})_{quantum} = K_{ontic}/(K_{inst})_{ontic}$ (the last equality is true by assumption of the theorem). We then simply have $K_{ontic}$ (the number of distinct quantum states) = $\infty$ and $(K_{inst})_{ontic} = N^2$ and thus $\gamma_{quantum} = \infty$, for such interpretations/theories. The second part of the theorem can be proved by simply noticing that, if quantum states themselves are either the sole ontology or only part of it, the set of ontic states cannot have less element than that of the quantum states, which is already $\infty$. 

This theorem corrects the mistake of using a knowledge independent gamma. It deals with the situation where the states in our theory are at the same level, i.e. they are instrumental and are by interpretation or definition, ontic. It is the correct version of Hardy's theorem if we realise that the excess baggage factor is meaningful only when it is defined on a single definite level of knowledge.

This theorem, in contrast to Hardy's original theorem, does not discredit the ontic approach in its entirety. It concerns only ontic theories in which the quantum states are ontic states or in which the ontologies include the quantum states. For example, the theorem implies the following approaches to quantum theory are implausible:
\begin{description}

\item[(a)] \textit{purely interpretational de Broglie-Bohm's approach}:
Note that this concerns the \textit{purely interpretational} view of de Broglie-Bohm's approach, which means that the underlying particle positions can never be measured nor give rise to any measurable effect, and thus the quantum states are in principle the only instrumental states (thus this does not concern the pilot-wave approach as proposed by Valentini). In this approach the wavefunction is part of the ontology and the ontic states are infinite in number. 
\item[(b)] \textit{Everettian many-worlds approach}:
There are many versions of this but we restrict our concern with the most common version of it - there is only one wavefunction, that of the universe, and it is the only ontology. Then even if the dimension of its Hilbert space is finite, there will still be continually many possible wavefunctions, and thus infinite relative excess baggage.
\end{description}

\section{\label{sec:level1}Conclusion}

There are two main results in this paper, the first is that Hardy's argument that ontic approaches to quantum theory requires an infinitely more superfluous resources than that in classical theories is untenable. The second result is that his quantum ontological excess baggage does not constitute a valid criterion for determining the plausibility of the ontic approaches to quantum theory. The reason for these mistakes is the failure to recognize that the instrumental states need not be the quantum states in the face of our knowledge about the ontic states. The correct form of the criterion, as proposed in this paper, is found to rule out only certain approaches such as the Everettian interpretation and the purely interpretational version of the de Broglie-Bohm theory.

\end{document}